# DHICA ion-modified melanin based porous Si solar cell

M. Semenenko[1,2,3], T. Yu. Obukhova[2], S. O. Kravchenko[1], O. S. Pylypchuk[3], O. Yu. Ostapenko[3,4], T. O. Kuzmenko[3,4], M. V. Voitovych[1], S. Davidenko[5], Ye. S. Davidenko[5], S. S. Davidenko[5], and A. Sarikov[1,2,4]

[1]V Lashkaryov Institute of Semiconductor Physics of the National Academy of Sciences of Ukraine, 41 Nauky Avenue, 03028 Kyiv, Ukraine

[2]National Technical University of Ukraine "Igor Sikorsky Kyiv Polytechnic Institute", 16 Politehnichna Street, 03056 Kyiv, Ukraine

[3]Institute of Physics of the National Academy of Sciences of Ukraine, 46 Nauky Avenue, 03028 Kyiv, Ukraine

[4]Educational Scientific Institue of High Technologies, Taras Shevchenko National University of Kyiv, 4-g Hlushkova Avenue, 03022 Kyiv, Ukraine

[5]ABC Smart Solutions AG, Aegeristrasse 5 Zug, ZUG, 6300, Switzerland

E-mail: semandko@gmail.com

**Abstract**

This work studies photosensitivity and current-voltage characteristics under illumination of water-soluble eumelanin films based on DHICA tetramers and hybrid eumelanin/porous Si photovoltaic cells aimed at optimizing their characteristics. By using DMSO to dissolve melanin containing protonated carboxyl groups and to remove ammonium cations, it became possible to significantly reduce the ionic component of conductivity, thus improved the electron transport. It was found out that dissolution in DMSO provides a denser π-π stacking of DHICA tetramers, which led to a significant improvement in the photovoltaic cell parameters. In particular, the efficiency increased from 0.023 % to 4.4 % and the series resistance decreased from 119 Ω to 42.6 Ω. Modeling demonstrated that high-temperature annealing, which causes decarboxylation, leads to a structural rearrangement of tetramers from a Christmas tree-like configuration to a "toothed helix". At this, one of the planes partially straightens. This emphasizes the critical importance of the morphology of the organic layer for photogeneration of carriers in a heterojunction.

## Introduction

Melanins belong to organic pigments and are an important class of biomacromolecules widely represented in nature and present in all the organisms, from animals to plants, fungi, and microorganisms [1-3].

The optical, mechanical and electrical properties of eumelanin are widely studied [4]. In [5], the mechanisms of charge transfer in eumelanin granules obtained by cold pressing of eumelanin powders are investigated. After heat treatment, an increase in photocurrent and a decrease in activation energy were observed. This decrease was from 0.56 eV to 0.44 eV for natural melanin and from 0.62 eV to 0.51 eV for synthetic melanin, respectively. In [6], electrical conductivity of melanin granules in a vacuum chamber upon gradual adding water vapor was studied. The obtained dependence of the conductivity on water content indicated that melanin is an ion-electron conductor [3]. In [7], dark and light conductivities of deposited melanin molecules were studied. The dark conductivity increased with an increase in temperature. After heat treatment of synthetic melanin films at 400 K, a decrease in photocurrent and photoconductivity was observed [8]. The work [9] studied thin films of synthetic eumelanin as semiconductor channels in organic transistors. An n-type conductivity in eumelanin films was experimentally demonstrated, which evidences the capacity of the material to operate as an electronic conductor. The authors of [10] proposed a way to create a bulk heterojunction based on porous Si (por-Si) and synthetic DHI-eumelanin. The DHI-melanin is embedded into the pores of Si and forms a photosensitive interface capable of generating photocurrent under illumination at the wavelengths of 400–850 nm. It was found out that por-Si impregnated with eumelanin demonstrates higher photoelectric current generation efficiency in the long-wave spectral region compared to an unfilled porous matrix. The work [11] showed that melanin with halide perovskite forms a composite with a broad absorption range from UV to near-IR and high quantum yield of photothermal conversion (99.56 %). The first prototype of a photoelectrical conveter based on eumelanin films as a separate object of investigation was proposed in [12]. The organic-inorganic heterojunction was formed by dropping of an aqueous solution of melanin on the por-Si surface. The photoelectrical conveter demonstrated a high fill factor FF = 90 % and an efficiency of 0.03 %.

Despite a large number of studies, the model of melanin conductivity remains undefined, and the studies are not systematized. Depending on the sample type (tablets, films, granules, self-organized structures, etc.), hydration conditions, and measurement method, interpretations of photoconductivity behavior varied from amorphous semiconductor to ion-electron conductor. Recent studies [3, 6] confirm that contemporary models of melanin conductivity are increasingly oriented towards hybrid (electronic and ionic) conductivity mechanisms [9, 13, 14]. Other key points in the study of melanin, in particular the interpretation of optical transmission spectra and determination of the eumelanin band gap, as well as the relationship between the eumelanin structure, the degree of hydration and the conductivity, were discussed in [13]. Light absorption in organic materials should be described within the optical resonator model. In melanin, such a resonator is formed by conjugated chains, which are sequences of double and single carbon atom bonds. The second key factor is the hydration degree of melanin. Since the studies do not distinguish between eumelanin types (DHI or DHICA), the influence of OH groups in different molecular fragments on conductivity remains unclear. As shown in [13], to ensure the reproducibility of conductivity measurement results, the eumelanin structure should be identified with infrared spectroscopy, the level of film hydration should be controlled, and contributions of each OH group and their effects on conductivity should be separated.

The aim of this work is to find ways to optimize the structure of water-soluble eumelanin films for formation of electronic conduction channels and reduction of the concentration of ions in the organic eumelanin film to increase photogeneration in eumelanin-por-Si heterojunction.

**Experimental**

Eumelanin films were obtained by depositing aqueous solutions of eumelanin on a substrate followed by drying. An aqueous solution of eumelanin with a concentration of 10 mg/ml (1 % solution) in water was prepared using the method described in [13]. This solution was further diluted with distilled water to concentrations of 1 mg/ml (0.1 %), 5 mg/ml (0.5 %), and 8 mg/ml (0.8 %). To deposit the films on the substrates, the respetie solutions were additionally diluted by adding 1 μg and 5 μg of 100 % DMSO. The rationale for using an additional solvent is discussed below. Various types of substrates were used for film deposition. The first type Sital plates with a contact grid with oppositely directed pins, was used for measuring electrical characteristics [13]. The second type were glass substrates used for measuring transmission spectra, reflectance, conductivity, etc.

Capacitance-frequency characteristics were measured using an R&C LX200 complex in automatic mode in parallel equivalent circuit configurations for CpD and |Z|Phase modifications. The measurements were carried out in the frequency range from 4 Hz to 500 kHz. Frequency dependence of capacitance and dissipation coefficient ($D$) were recorded. The amplitude of the alternating signal was set at 0.1-1 V to ensure stable and accurate readings in the entire measurement range. To evaluate changes in the dielectric behavior of the samples, the values of capacitance and $D$ were studied in dark and under UV lamp illumination (< 400 nm, 780 Lm at a distance of 5 cm).

Current-voltage (*I-V*) characteristics were measured using an SMU-Keithley-2450 complex. The light *I-V* characteristics were measured under controlled AM1.5 conditions at a light source power of 100 mW/cm². The spectral characteristic was monitored using digital RGBC sensors and a CCD diffraction spectrometer. The light intensity was monitored by a UNi-T UT381A luxmeter using a Plexiglas thermal filter [15]. The photoelectric converter was manufactured in accordance with [12]. From bottom to top it was: Al/bulk p-Si KDB-1/1-3 μm por-Si/10 nm SiO$_2$/DHICA/65 nm ITO/Ag grid.

ChemOffice, Avogadro, ChimeraX, and Gaussian_09W software were used to model oligomer molecules and their agglomerates. Energy surfaces were obtained through structure optimization by minimizing the total potential energy of the molecule (MM2 minimization) using Chem3D. The HOMO and LUMO values of the energy levels were obtained using the Extended Huckel algorithm built into Chem3D.

**Results and discussions**

1 mg/ml (0.1 %), 5 mg/ml (0.5 %) and 8 mg/ml (0.8 %) aqueous solutions of eumelanin with and without addition of 1 μg and 5 μg of 100 % DMSO were deposited on the surface of the substrates and dried at room temperature. When drops are deposited on the substrates, both water and DMSO evaporate during drying, forming a homogeneous eumelanin film. Unlike pure DHI or DHICA material, the melanin films obtained in our experiments were water-soluble. The reason for water solubility is that melanin is extracted from plants with adding an aqueous ammonium solution, which ionizes carboxyl groups in DHICA to form the corresponding carboxyl anion. The parent molecules DHI and DHICA are water-insoluble. When ionized, DHI mainly precipitates. We believe therefore that water-soluble eumelanin films are concentrated DHICA agglomerated complexes. DMSO, as a highly polar aprotic solvent that does not form hydrogen bonds, effectively separates the polymer composition diluted in water and allows the film to polymerize again upon drying.

Frequency dependence of capacitance and $D(\omega)$ were recorded in the CpD and |Z|Phase configurations. At high frequencies, a sharp increase in capacitance was observed for all the DHICA samples without DMSO. At this, there were no noticeable differences between UV-illuminated and non-illuminated samples. This means that electronic polarization did not undergo noticeable changes upon UV illumination. $D(\omega)$ shows a decrease in dielectric losses with frequency up to 70 kHz, which is characteristic of a parallel CpD substitution circuit. As a rule, this case characterizes losses caused by electrical conductivity. The frequency dependences of capacitance and $D$ do not demonstrate relaxation or oscillatory nature of the scattering mechanism. Therefore, the respective graphs are not presented in this work. However, we analyze them to interpret the impedance measurement results. Impedance was measured in the |Z|Phase configuration. The results for the DHICA melanin films with DMSO without illumination are presented in Fig. 1 as a Nyquist diagram. The Nyquist diagram [16] is constructed using expressions (1) and shows the relationship between the resistive and reactive impedance components in different frequency ranges of a parallel

equivalent circuit.

$$Z_{\Re} = \frac{1}{\omega C_p} \frac{D}{1+D^2}, Z_{\Im} = \frac{1}{\omega C_p} \frac{1}{1+D^2} \qquad (1)$$

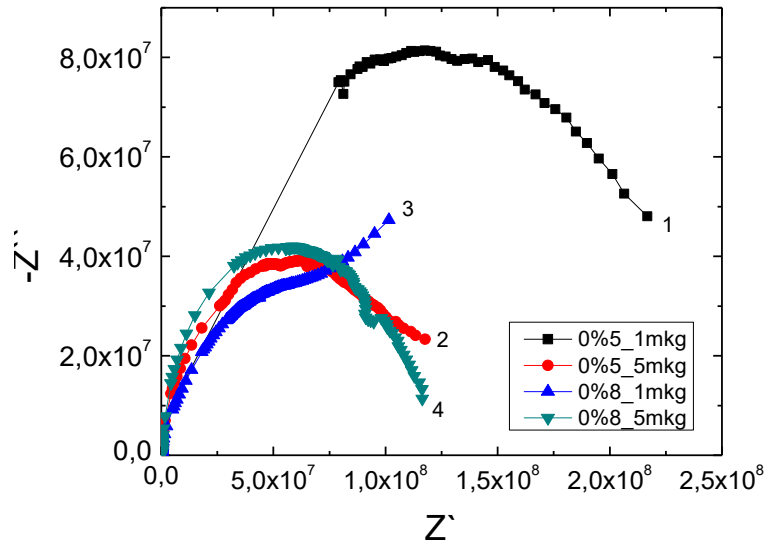

Figure 1. Nyquist diagram of the samples without illumination: 1 — 0.5 % + 1 μg DMSO, 2 — 0.5 % + 5 μg DMSO, 3 — 0.8 % + 1 μg DMSO, and 4 — 0.8 % + 5 μg DMSO.

As can be seen from Fig. 1, the diagrams of all the samples in the high-frequency region (closer to zero) are arcs of asymmetric semicircles. This fact indicates that at these frequencies, the equivalent circuit of the sample is a parallel RC or R-CPE chain (a constant phase element, i.e., the capacitance is modeled by a parasitic shunt and/or series resistor), which is consistent with the monotonic frequency decline of $D(\omega)$ (not shown here). Increase in the semicircle radius for the 0.5 % sample with added 1 μg of DMSO (curve 1) indicates an increase in impedance, which may point to a decrease in conductivity or weakening of internal polarization. This is due to the influence of the DMSO as a solvent that separates eumelanin and reduces the conductivity efficiency by dissociating aggregated oligomers with a corresponding decrease in the electronic component of the conductivity. A distinct feature of the 0.8 % 1 μg sample (curve 3) is the presence of a linear section with a slope of ~45° to the real $Z`$ axis in the low-frequency region, indicating the dominance of conductivity over polarization. This type of impedance is characteristic of materials with dominant ionic conductivity and partially blocking electrodes.

The frequency dependence of the absolute AC conductivity $G(\omega)$ is traditionally analyzed using Jonscher's projection (Jonscher's Universal Power Law) [17], which is described by the power function $G(\omega) = G_0 + A\omega^s$, where $G_0$ and $A$ are coefficients, and $s$ is the slope of the curve in coordinates log(*G*) *vs.* log(ω). Hence, the absolute value of the conductivity was obtained directly from complex impedance measurements using the following expression:

$$G = \frac{\cos(\varphi)}{|Z|} \qquad (2)$$

where *φ* is the impedance phase angle in radians and |*Z*| is the absolute value of the impedance in Ω.

Fig. 2 shows the frequency dependence of the absolute conductivity $G(\omega)$ (1/Ω). At low frequencies (10–10$^3$ rad/s), the values of *s* are very small and localized relaxation prevails. This indicates dominance of the stationary (DC) component and slow processes (e.g., interphase or ionic conductivity). In the mid-frequency range (10$^3$–10$^4$ rad/s), a *s* smoothly increases, which corresponds to transition to a different conductivity mechanism. At high frequencies (10$^4$–10$^5$ rad/s), an ionic jump takes place. The value of *s* ~ 1, i.e. the transport processes become faster, possibly involving more localized carriers [18-20].

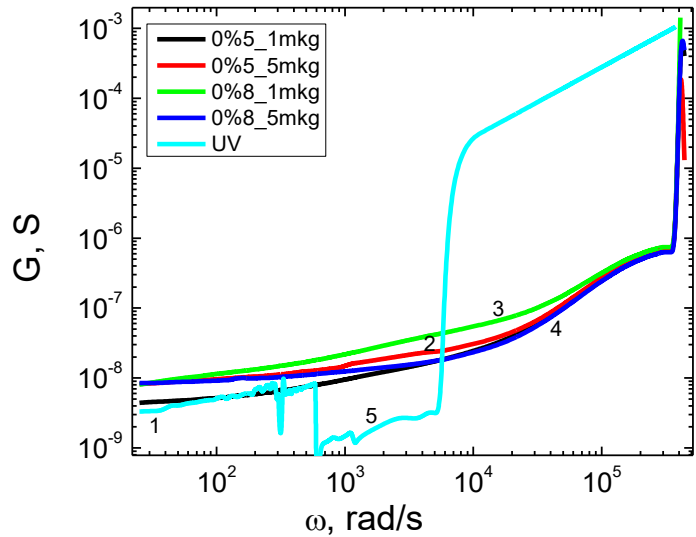

Figure 2. AC conductivity $G(\omega)$ of eumelanin films on Sital substrate: 1 — 0.5 % + 1 μg DMSO, 2 — 0.5 % + 5 μg DMSO, 3 — 0.8 % + 1 μg DMSO, and 4 — 0.8 % + 5 μg DMSO.

At low frequencies, a decrease in conductivity of the 0.5 % 1 μg sample (curve 1) compared to other samples is observed. This indicates that the ionic charge transfer mechanism is less pronounced or the interphase polarization effect is absent in this sample.

The conductivity of the 0.8 % 1 μg sample (curve 3) is the highest. The slopes of the Jonscher's curve for this sample are as follows: the range (4-10$^2$ ) *s* = 0.25, the range (10$^2$-10$^3$ ) *s* =

0.28, the range ($10^3$-$10^4$) $s = 0.4$, the range ($10^4$-$10^5$) $s = 0.81$, and the range ($10^5$ and higher) $s = 0.77$. The analysis of the slopes is presented below.

In the low-frequency range, $s \approx 0.25$-$0.28$. Very low values of $s$ are a classic manifestation of electrode/interphase polarization and ion migration with charge accumulation at contacts (Maxwell–Wagner–Sillars, "blocking electrodes"). The Nyquist curve at $f < \sim 100$ Hz being almost a straight line (mainly Re(Z)) corresponds to an ohmic channel with an admixture of slow polarization. The conductivity is mainly ionic in this range. The electronic contribution is small and is "smothered" by the space charge at the contacts. Physically, this is not "electronic ohmic" conductivity, but a consequence of very slow processes such as ion migration and polarization at the contacts, i.e., the Maxwell–Wagner effect. Therefore, this section still corresponds to AC behavior but with such a small $s$ value that it looks almost horizontal. For a heterojunction in a photoelectric converter, such behavior is undesirable because it means that slow charge drift can slow down the converter response to changes in illumination and cause hysteresis. An energy will be lost to recharging ions and interface layers. This will reduce the fill factor and stability even if the instantaneous electron conductivity at higher frequencies is good.

In the mid-range, $s \approx 0.40$. This corresponds to a transition to dispersed ion jump conductivity. The ions no longer have time to form a full spatial charge during the field period, and increasing AC conductivity appears. In the high-frequency range, $s \approx 0.81$ and then $s \approx 0.77$. This corresponds to tunneling between π-stacks or fast localized jump electron conductivity within melanin domains. The electrode polarization no longer dominates. Electronic paths along partially stacked areas become manifested. At low frequencies (up to ~500 rad/s), the conductivity of the illuminated sample (curve 5) is reduced, but the slope does not significantly increase, pointing to photo-stimulated release of charge carriers in near-electrode regions.

The conclusion about the frequency dependence of the conductivity is as follows: The conductivity in the range up to ~$10^3$ rad/s is mainly defined by migration of dissolved and residual ions ($NH^{4+}$, $H^+/OH^-$) and interphase polarization at the indium/film/ceramic contacts. For stable electronic contact and dominant electronic transport through eumelanin, a set of approaches to suppress ion mobility and reduce electrode polarization should be developed. In the "useful" low-frequency window, the conductivity is ionic and contact-polarization. The electronic channel becomes noticeable only at higher frequencies, where $s \approx 0.8$. For photoelectric converters, presence of mobile ions and water molecules must be technologically limited, and chemically stable electrodes must be used. The following approaches for modifying ion-electron conductivity with electrode polarization into electron-ion one are proposed:

1. Replacement of the ammonium cation $NH^{4+}$ with a bulky, immobile cation or protonation of the carboxyl group followed by drying and heat treatment.

2. Drying in a vacuum/inert gas atmosphere, controlling humidity during measurements.

3. Use of inert Au/Pt electrodes to distinguish injection from In from pure ionic polarization.

The following ways may be suggested to improve the photoconductivity of melanin films. We look first at the formation of eumelanin film components. Dissociation-association of two melanin oligomers, each of which consists of e.g. four DHICA monomers, is realized in an aqueous medium through acid-base regulation of the solution. This leads to a change in the solution pH as schematically represented in Fig. 3. Acidification of the solution leads to precipitation of a black precipitate due to protonation of carboxyl groups in DHICA with formation of hydrogen bonds between the oligomers, which causes their aggregation with corresponding partial $\pi$ stacking between them (see the lower part of the diagram). On the other hand, alkalization of the solution leads to dissolution of the precipitate due to ionization of carboxyl groups in DHICA with formation of a negative charge on them and corresponding anions. This causes repulsion between the oligomers, their dissociation and disappearance of $\pi$ stacking between them (see the upper part of the diagram in Fig. 3).

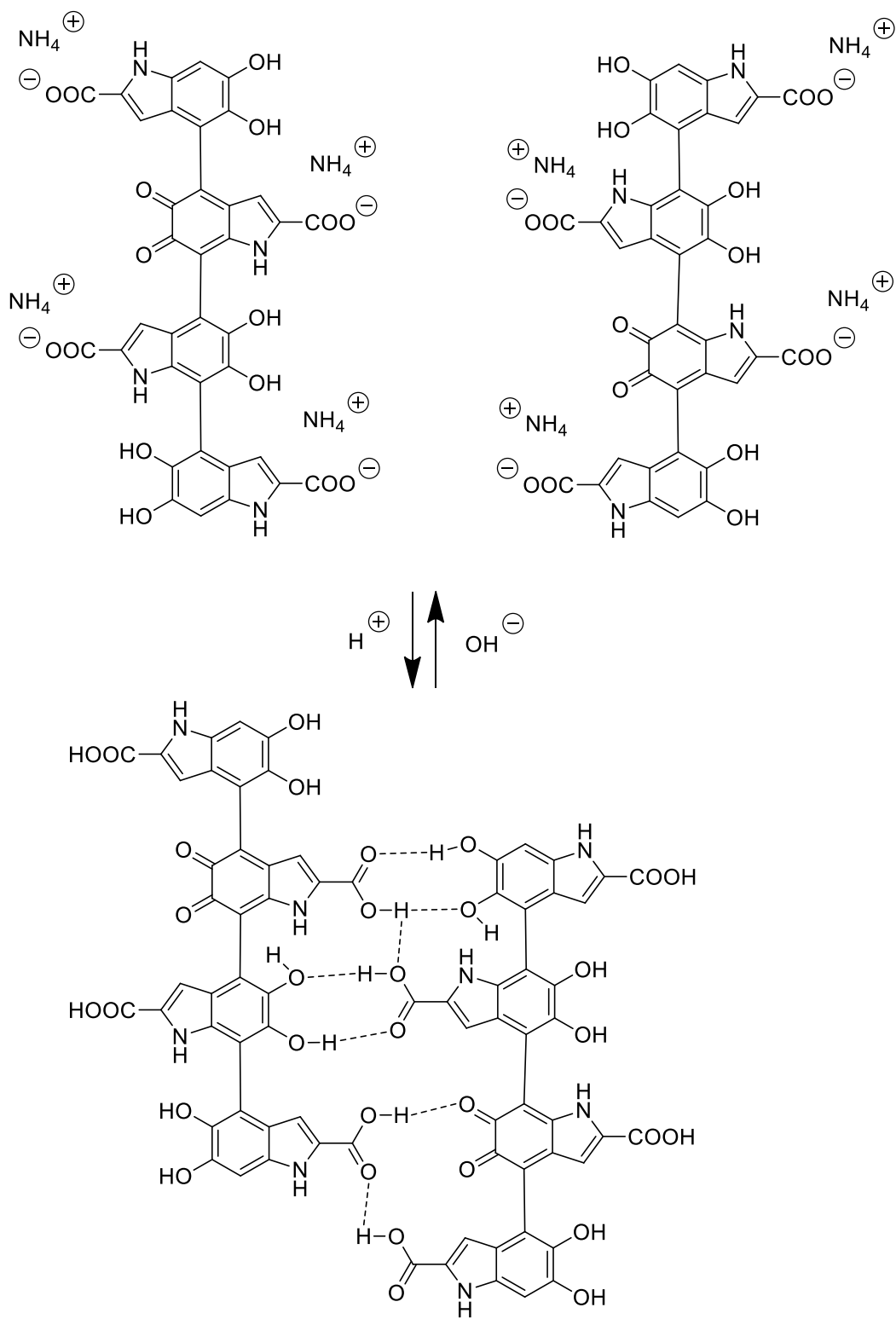

Figure 3. Schematic representation of the aggregation mechanism of eumelanin oligomers containing DHICA.

Appearance of black sediment after adding HCl is a typical indicator of more developed π-conjugation of eumelanin molecule chains. That is, due to appearance of hydrogen bonds in an acidic environment, a partially ordered network forms by convergence of DHICA monomers with a respective increase in the conjugated chain lengths (see Fig. 3, lower part of the diagram). Filtering the the precipitate leads to removal of ammonium ($NH^{4+}$) cations, which decreases the ionic component in the melanin film after drying. The transmission spectra of $DHICA^+NH^{4+}$, which are present in the many works, are characteristic. The transmission spectra of DHICA or DHI are monotonous. Therefore, visual evaluation of transmission spectra enable obtaining information about the degree of agglomeration of melanin oligomers and presence of ions. When dissolving the DHICA precipitate obtained by acidifying the solution, hydrogen bonds between the monomers of individual melanin oligomers in DMSO break, resulting in oligomer partial dissociation. However, after evaporation of DMSO, denser surface structures compared to the films obtained after drying aqueous melanin solutions in presence of ammonium (aqueous solution of $DHICA^+NH^{4+}$) are observed.

We consider the effect of agglomeration of melanin oligomers without free ions on conductivity with an emphasis on electron conductivity. Association between neighboring melanin oligomers resulting from acidification of the solution causes better contact between π-domains, increasing the DC component of the absolute conductivity $G_0$ and decreasing the number of deep traps at the "pancakes" interfaces. The ionic conductivity of the film decreases after removal of $NH_4^+$ cations from the bulk. But presence of a carboxyl group (COOH) may lead to surface monomolecular adsorption of water molecules due to formation of hydrogen bonds. This dampens accumulation of spatial charge at the contacts and reduces hysteresis. As a result, low-frequency behavior becomes more stable and relaxation times become shorter. Hence, dissolution of melanin with protonated carboxyl groups in DMSO improves the electron component of conductivity due to ordered dissociation of oligomers, which, after precipitation on the surface, allows them to pack more tightly due to the reformation of hydrogen bonds leading to an increase in the degree of π-π stacking between neighboring melanin oligomers. However, the final conductivity will also depend on the extent to which DMSO can modify the functional groups of melanin [21]. This work reported that DMSO can lead to sulfonation of hydroxyl groups, i.e., replace hydrogen with methylsulfonyl ($SO_2CH_3$), causing a change in the local electronic configuration. This is sometimes useful because it reduces the proportion of proton conductivity. However, excessive sulfonation can weaken aggregation between melanin oligomers. Therefore, the expectations are as follows: a smaller low-frequency "tail" on the Nyquist diagram, higher $G_0$ and $s$ values in the mid-high-frequency range, and, in general, a shift of the balance towards the electron channel. The absolute conductivity increases if sulfonation does not interfere with formation of effective π-chains.

Having clarified the formation mechanism of water-soluble eumelanin, we proceed to evaluation of the energy and structural characteristics of the eumelanin films. IR spectroscopy [13] showed that carboxyl groups are present in water-soluble melanin. This indicates that eumelanin contains DHICA monomers that form a minimal structure in the form of tetramers upon polymerization. These tetramers agglomerate into larger complexes upon drying. To compare different shapes and complex structures of eumelanin molecules, we use ChemOffice and Avogadro software packages. Figure 4 shows models of main eumelanin monomers DHICA and DHI and the calculated values of the main LUMO-HOMO levels.

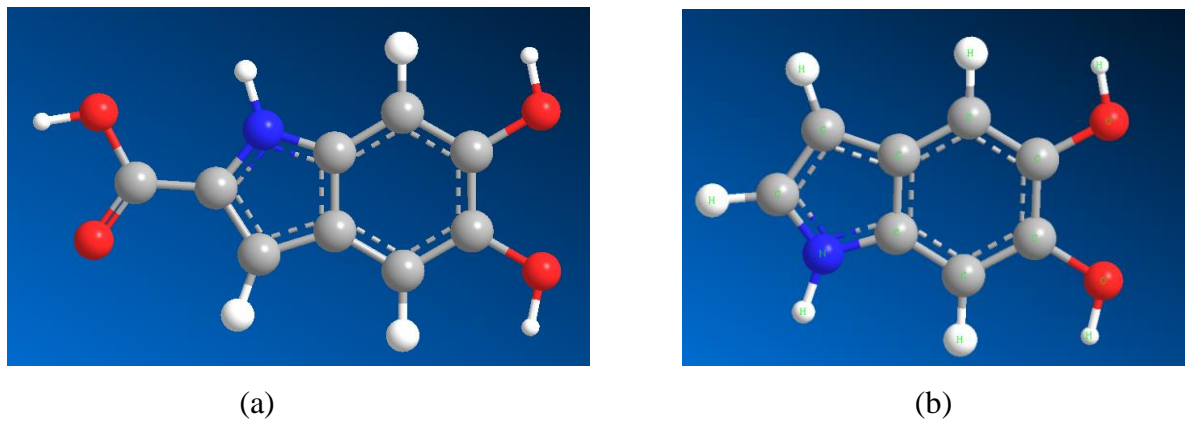

(a) (b)

Fig. 4. Melanin monomers DHICA and DHI and calculated values of the main LUMO-HOMO levels. (a) – DHICA monomer: LUMO = -2.7 eV and HOMO = -10.8 eV. (b) DHI monomer: LUMO = +0.4 eV and HOMO = -11 eV.

Upon agglomeration, the monomers shown in Fig. 4 form structures ranging from tetramers to larger agglomerates. We assume that the minimum unit that can dissolve in DMSO are tetramers, the images of which are shown in Fig. 5. The DHICA and DHI tetramers differ in the planarity of their structure. The DHI tetramers orient themselves into a planar structure during MM2 energy minimization (Fig. 5(c) and (d)), while the DHICA tetramers form a tree-like configurations (Fig. 5(a) and (b)). The planarity of the structure promotes better adhesion to the Si surface, as shown in [22] for monomers and agglomerated complexes of DHI molecules (Fig. 5(d)).

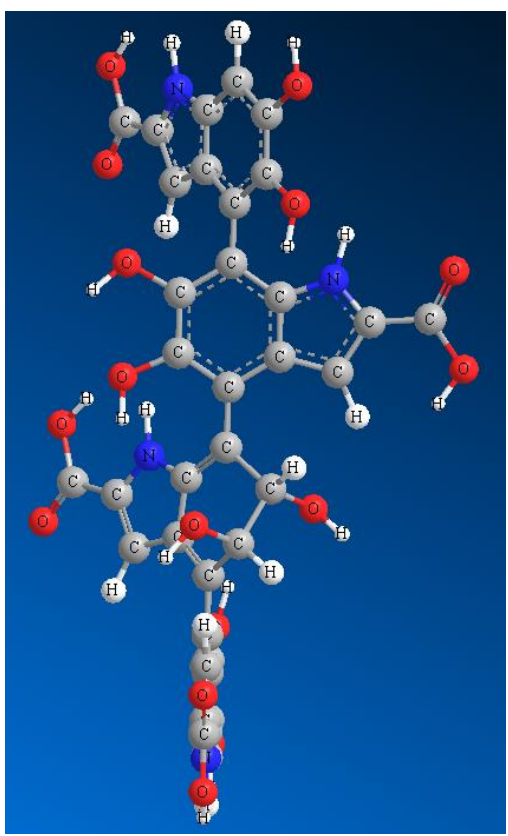
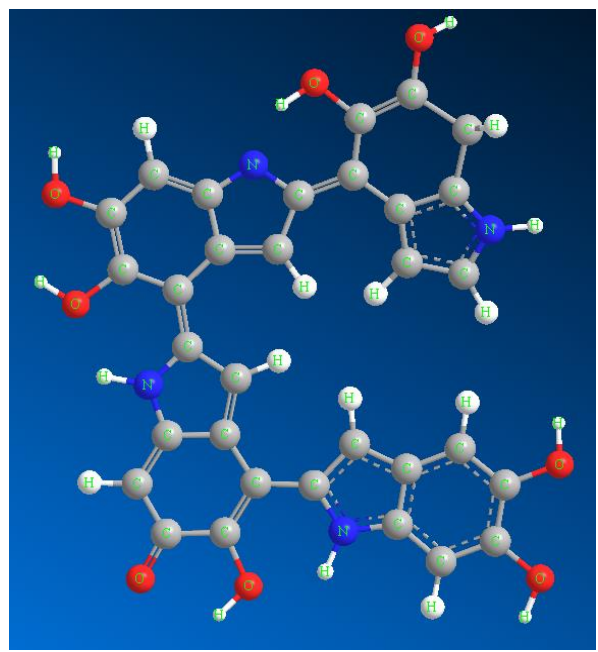

(a)

(c)

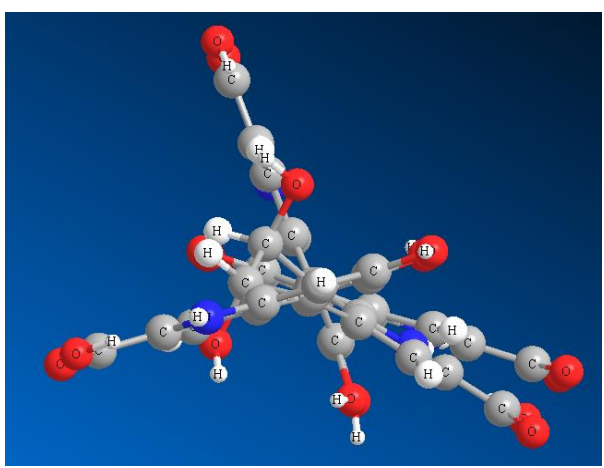
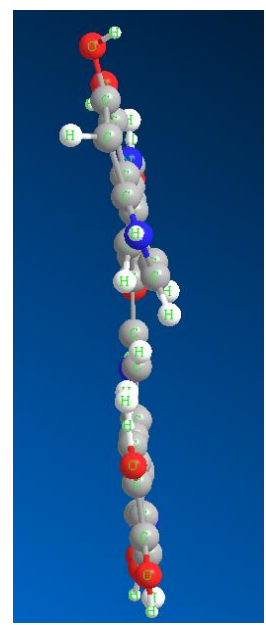

(b)

(d)

Fig. 5. DHICA (a, b) and DHI (c, d) tetramers and corresponding LUMO and HOMO values. (a, b) - DHICA tetramer: LUMO = -5.24 eV and HOMO = -7.1 eV. (c, d) - DHI tetramer: LUMO = (-1.7, -3.6, and -7.5) eV and HOMO = -8.9 eV.

We carried out modeling of temperature annealing for de-carboxylation procedure, which is equivalent to annealing in vacuum at a temperature of 400 to 600°C. In the simulation, we assumed that eumelanin tetramers and larger DHICA agglomerates lose their carboxyl group and water and sinter into water-insoluble complexes. When the carboxyl group is removed, the vacant site in the indole ring of the DHICA becomes active, and the agglomerated complexes should rearrange to align the tetramer plane. As can be seen in Fig. 6, when the structure energy is minimized, the tree-like shape of the DHICA tetramer (Fig. 6(a)) transforms into a toothed spiral (Fig. 6(b), (c) and (d)). That is, vacuum annealing of water-soluble eumelanin films, which removes carboxyl groups and water and changes the plane, does not provide a sufficient plane, unlike DHI tetramer, which is flat already in its initial state.

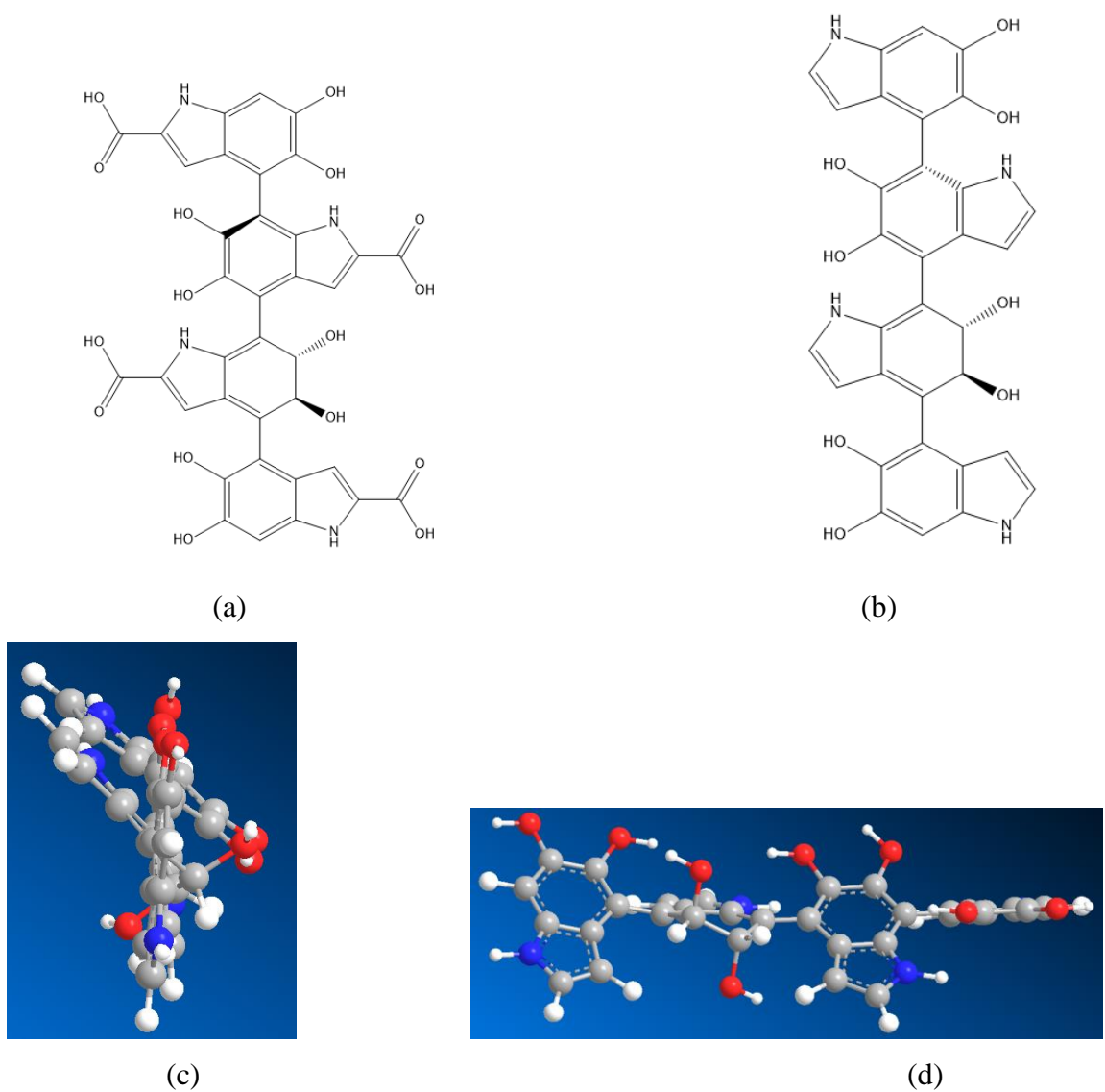

Fig. 6. De-carboxylation procedure, equivalent to vacuum annealing at temperatures ranging from 400°C to 600°C. (a) and (c) — initial DHICA tetramer, (b) and (d) - DHICA(modified) after annealing of DHICA(modified) tetramer: LUMO = -3.8 and HOMO = -7.1 eV.

Light *I-V* characteristic of the photoelectric converter fabricated according to [12] is shown in Fig. 7(a). After treating the aqueous solution in HCl and drying the resulting precipitate, the dry DHICA material without ammonium ($NH_4^+$) cations was dissolved in DMSO and deposited onto the por-Si surfaces. The *I-V* curve of the DHICA without cations is shown in Fig. 7(b) (curve 1). For comparison, the curve 2 is the *I-V* characteristic of the obtained from water-soluble DHICA films, i.e., in presence of cations, but the aqueous solution of eumelanin was diluted with DMSO when deposited. The experimental data were approximated using a two-diode model as described in [15, 23]. When dissolved in DMSO, the DHICA film (aqueous solution of DHICA+$NH^{4+}$) separates into tetramers and, when drying on the surface of the por-Si, packs more densely stacks better. The parameters of the por-Si photoelectric converter with the DHICA films are presented in Table 1. Due to the better stacking of eumelanin molecules, the efficiency of the converter increased from 0.023 % to 4.4 % and the series resistance decreased from 119 Ω to 42.6 Ω.

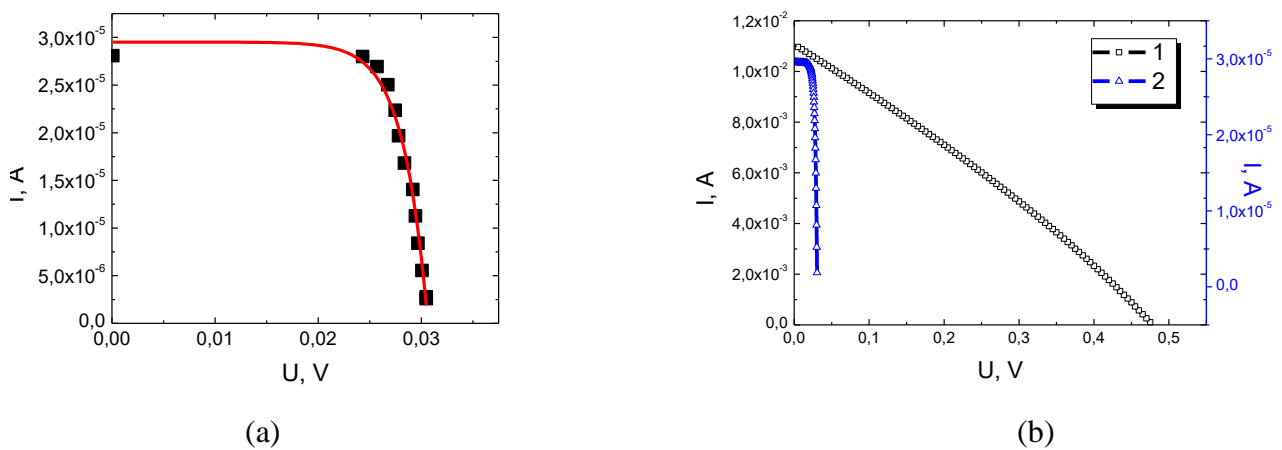

(a) (b)

Fig. 7. *I-V* characteristic of a photoelectric conveter with the following structure (from bottom to top): Al / bulk p-Si KDB-1 / 1-3 μm por-Si / 10 nm $SiO_2$ / DHICA tetramer / 65 nm ITO / Ag grid. (a) DHICA tetramer in water solution, (b): 1 – DHICA(modified) tetramer in DMSO ($A = 0.34$ cm$^2$), 2 - DHICA tetramer in water solution.

Table 1. Specific parameters of a single photoelectric converter with por-Si and DHICA films

|  | $N$ | $U_{oc}$, mV | $I_{sc}$, mA | FF | Ef, % | $R_s$, Ω | $R_{sh}$, Ω |
|---|---|---|---|---|---|---|---|
| Water-soluble film | a | 31 | 0.03 | 0.732 | 0.023 | 119 | 6042 |
| DHICA(modified) tetramer in DMSO | b1 | 475 | 10 | 0.287 | 4.4 | 42.6 | 54.3 |

The band diagram of the photoelectric converter structure was modeled with reference to the works on electronic field emission from por-Si resonant-tunnel structures [24, 25] and linking the diameter of Si crystallites to the photoluminescence band [26, 27]. It was assumed that photoluminescence and resonant field emission are observed simultaneously at the minimum possible sizes of por-Si crystallites, for which electron transport limitations due to increase in the bandgap width begin to manifest themselves. For the initial p-Si, the electron affinity is -4.15 eV (the bandgap width = 1.17 eV). For por-Si crystals with diameters of at least 2.8 nm, the electron affinity will be equal to 3.65 eV [24, 25]. The bandgap value calculated using an empirical expression [24] is about 1.81 eV. The Si oxide should be as thin as possible to enable injection of holes and electrons during *I-V* measurements. The value of the $SiO_2$ electron affinity is 1 eV, and its bandgap width is 9 eV. The key values of the band diagram of the photoelectric converter structure based on water-soluble agglomerates of DHICA tetramers and por-Si are presented in Table 2.

Table 2. Band diagram of the photoelectric converter structure (from bottom to top): Al / bulk p-Si KDB-1 / 1-3 μm por-Si / 10 nm $SiO_2$ / DHI-DHICA monomers-tetramers / 65 nm ITO /Ag grid.

|  | ITO | DHI-t | DHI-m | DHICA-mod | DHICA-t | DHICA-m | $SiO_2$ | por-Si | p-Si | Al |
|---|---|---|---|---|---|---|---|---|---|---|
|  |  |  |  |  |  |  |  |  |  |  |
| Electron affinity ($\chi$, $E_f$), eV | -5.2 |  |  |  |  |  | -1 | -3.7 | -4.2 | -4.4 |
| Bandgap ($E_g$), eV |  |  |  |  |  |  | 9 | 1.8 | 1.2 |  |
| LUMO ($E_c$), eV |  | -7.5 | +0.4 | -3.8 | -5.2 | -2.7 | -1 | -3.7 | -4.2 |  |
| HOMO ($E_v$), eV |  | -8.9 | -11 | -7.1 | -7.1 | -10.8 | -10 | -5.5 | -5.3 |  |

**Conclusion**

It is found out that the key factor for improving the efficiency of hybrid photoelectric converters based on eumelanin-por-Si of the configuration (from bottom to top) Al/bulk p-Si KDB-1/1-3 μm por-Si/10 nm $SiO_2$/DHI-DHICA monomers-tetramers/65 nm ITO/Ag grid is control of the morphology of the organic layer and minimization of its ionic conductivity. Use of DMSO as a solvent for DHICA tetramers as well as filtration procedure to remove $NH^{4+}$ cations successfully reduce the ionic component of the conductivity and ensure denser π-π stacking of oligomers. Dissolution of eumelanin with protonated carboxyl groups in DMSO improves the electronic component of the conductivity due to ordered dissociation of oligomers. This allows molecules to pack more tightly after precipitation on the surface due to the reformation of hydrogen bonds with corresponding increase in the degree of π-π stacking between neighboring melanin oligomers. Such structural optimization, favoring formation of efficient electronic channels, leads to a significant increase in the efficiency of the photoelectric converter from 0.023 % to 4.4 % and a decrease in the series resistance from 119 Ω to 42.6 Ω. Vacuum annealing simulation (decarboxylation at 400-600°C) confirms that thermal methods transform the initial water-soluble DHICA tree-like structure into a spiral DHICA-modified structure with partial alignment of one plane, which is an improvement over the initial structure but does not provide the desired completely flat morphology (which is characteristic of water-insoluble DHI tetramers). Therefore, another type of melanin that provides a flat structure in its initial state should be considered to further improve the efficiency of photoelectric converters.

**Authors' roles**

M. Semenenko: conceptualization, formulation of research tasks, analysis of the results; T. Yu. Obukhova: experimental setup for solar cells, sample preparation, data processing and analysis; S. O. Kravchenko: development of technological routes for formation of DHICA solutions, modeling; O. S. Pylypchuk: measuring and analyzing capacitance-frequency and impedance characteristics; O. Yu. Ostapenko: photo-conductivity measurements; T. O. Kuzmenko: preparation of DHICA based solutions; M. V. Voitovych: data processing and analysis; S. Davidenko, Ye. Davidenko, and S. S. Davidenko - melanin production; A. Sarikov: coordination the research, discussion of the results and preparation of the publication.

**Acknowledgement**

M. O. Semenenko, O. Yu. Ostapenko, M. V. Voitovych and A. Sarikov acknowledge support of their contribution by the project 4Ф-2024 "Multilayer structures with organic polymer semiconductor heterojunctions and Si photon crystal based backside reflectors for photoelectric solar energy converters" of the Program of Joint Research Projects of Scientists of the Taras Shevchenko National University of Kyiv and the National Academy of Sciences of Ukraine in 2024-2025. O. S. Pylypchuk, O. Yu. Ostapenko, T. O. Kuzmenko and M. O. Semenenko acknowledge support by the Project No. 5.8/25-П "Energy-saving and environmentally friendly nanoscale ferroics for the development of sensorics, nanoelectronics and spintronics" of the Target Program of the National Academy of Sciences of Ukraine. T. Yu. Obukhova, A. Sarikov and M. O.

Semenenko also acknowledge support of their research by the Project "Development of photovoltaic cells and thin-film photoconverters based on organic solar light absorbers" of the Ministry of Education and Science of Ukraine, state registration number 0125U001660.**References**

1. A. Bernardus Mostert, Melanin, the what, the why and the how: An introductory review for materials scientists interested in flexible and versatile polymers, Polymers **10**, 1670 (2021). https://doi.org/10.3390/polym13101670.

2. S. Singh, S. B. Nimse, D. E. Mathew, A. Dhimmar, H. Sahastrabudhe, A. Gajjar, V. A. Ghadge, P. Kumar, P. B. Shinde, Microbial melanin: Recent advances in biosynthesis, extraction, characterization, Biotechnol. Adv. **53**, 107773 (2021). https://doi.org/10.1016/j.biotechadv.2021.107773.

3. A. B. Mostert, B. J. Powell, F. L. Pratt, G. R. Hanson, T. Sarna, I. R. Gentle, P. Meredith, Role of semiconductivity and ion transport in the electrical conduction of melanin, PNAS **109**, 8943 (2012). http://www.pnas.org/cgi/doi/10.1073/pnas.1119948109.

4. M. Brinza, J. Willekens, M. L. Benkhedir, E. V. Emelianova, G. J. Adriaenssens, Photoconductivity methods in materials research, J. Mater. Sci.: Mater. In Electronics **16**, 703 (2005). https://doi.org/10.1007/s10854-005-4972-7.

5. T. Ligonzo, M. Ambrico, V. Augelli, G. Perna, L. Schiavulli, M. A. Tamma, P. F. Biagi, A. Minafra, V. Capozzi, Electrical and optical properties of natural and synthetic melanin biopolymer, J. Non-Cryst. Solids **355**, 1221 (2009). https://doi.org/10.1016/j.jnoncrysol.2009.05.014.

6. A. Bernardus Mostert, B. J. Powell, I. R. Gentle, P. Meredith On the origin of electrical conductivity in the bio-electronic material melanin, Appl. Phys. Lett. **100**, 093701 (2012). http://dx.doi.org/10.1063/1.3688491.

7. M. M. Jastrzebska, K. Stępień, J. Wilczok, M. Porebska-Budny, T. Wilczok, Semiconductor properties of melanins prepared from catecholamines, Gen. Physiol. Biophys. **9**, 373 (1990).

8. P. R. Crippa, V. Cristofoletti, N. Romeo, A band model for melanin deduced from optical absorption and photoconductivity experiments, Biochim. et Biochim. Acta **538**, 164 (1978). https://doi.org/10.1016/0304-4165(78)90260-X.

9. N. L. Nozella, J. V. Paulin, G. L. Nogueira, N. B. Guerra, R. F. de Oliveira, C. F. O. Graeff, Probing n-type conduction in eumelanin using organic electrochemical transistors, ACS Appl. Electron. Mater. **7**, 3176 (2025). https://doi.org/10.1021/acsaelm.5c00293.

10. G. Mula, L. Manca, S. Setzu, A. Pezzella, Photovoltaic properties of PSi impregnated with eumelanin, Nanoscale Res. Lett. **7**, 377 (2012). https://doi.org/10.1186/1556-276X-7-377.

11. K. Wang, Y. Hou, B. Poudel, D. Yang, Y. Jiang, M.-G. Kang, K. Wang, C. Wu, S. Priya,

Melanin–perovskite composites for photothermal conversion, Adv. Energy Mater. **9**, 1901753 (2019). https://doi.org/10.1002/aenm.201901753.

12. D. Volynskyi, M. Dusheiko, R. Madan, N. Kutuzov, T. Obukhova, Melanin/porous silicon heterojunctions for solar cells and sensors applications, 2020 IEEE 40th Int. Conf. on Electronics and Nanotechnol. (ELNANO) 343 (2020). https://doi.org/10.1109/ELNANO50318.2020.9088805.

13. T. Obukhova, M. Semenenko, M. Dusheiko, S. Davidenko, Ye. S. Davidenko, S. S. Davidenko, S. Malyuta, S. Shahan, O. S. Pylypchuk, P. Mochuk, M. Voitovych, T. Kuzmenko, A. Sarikov, Optical studies of melanin films as a material for solar light absorbers, Mater. Res. Express **12**, 065503 (2025). https://doi.org/10.1088/2053-1591/ade227.

14. N. B. Guerra, J. V. M. Lima, N. L. Nozella, M. H. Boratto, J. V. Paulin, C. F. de Oliveira Graeff, Electrochemical doping effect on the conductivity of melanin-inspired materials, ACS Appl. Bio Mater. **7**, 2186 (2024). https://doi.org/10.1021/acsabm.3c01166.

15. M. Semenenko, M. Dusheiko, G. Okrepka, R. Redko, S. Antonin, V. Hladkovskyi, V. Shvalagin, F. Gao, S. Shahan, A. Sarikov, Vertically-aligned p-n junction Si solar cells with CdTe/CdS luminescent solar converters, Thin Solid Films **761**, 139536 (2022). https://doi.org/10.1016/j.tsf.2022.139536.

16. J. Wünsche, Y. Deng, P. Kumar, E. Di Mauro, E. Josberger, J. Sayago, A. Pezzella, F. Soavi, F. Cicoira, M. Rolandi, C. Santato, Protonic and electronic transport in hydrated thin films of the gigment eumelanin, Chem. Mater. **27**, 436 (2014). https://doi.org/10.1021/cm502939r.

17. P. P, S. Kumar, A. K. Gathania, A. K. Singh, Supreet, J. Prakash, S. Kumar, P. Malik, R. Castagna, G. Singh, Effect of carbon dots on tuning molecular alignment, dielectric and electrical properties of a smectogenic cyanobiphenyl-based liquid crystal material, J. Phys. D: Appl. Phys. **57**, 355302 (2024). https://doi.org/10.1088/1361-6463/ad4a84.

18. M. R. Díaz-Guillén, J. A. Díaz-Guillén, A. F. Fuentes, J. Santamaría, C. León, Crossover to nearly constant loss in ac conductivity of highly disordered pyrochlore-type ionic conductors, Phys. Rev. B **82**, 174304 (2010). https://doi.org/10.1103/PhysRevB.82.174304.

19. S. R. Elliott, Physics of amorphous materials. London, New York: Longman Group LTD, 1984. https://doi.org/10.1002/crat.2170200922.

20. A. Mansingh, AC conductivity of amorphous semiconductors, Bull. Mater. Sci. **2**, 325 (1980). https://www.ias.ac.in/article/fulltext/boms/002/05/0325-0351.

21. J. Wünsche, F. Cicoira, C. F. O. Graeff, C. Santato, Eumelanin thin films solution: processing growth and charge transport properties, J. Mater. Chem. B **1**, 3836 (2013). https://doi.org/10.1039/C3TB20630K.

22. A. Antidormi, C. Melis, E. Canadell, L. Colombo, Assessing the performance of eumelanin/Si interface for photovoltaic applications, J. Phys. Chem. C **121**, 11576 (2017).

https://doi.org/10.1021/acs.jpcc.7b02970.

23. M. O. Semenenko, M. G. Dusheiko, S. V. Mamykin, V. O. Ganus, M. V. Kirichenko, R. V. Zaitsev, M. M. Kharchenko, N. I. Klyui, Effect of plasma, RF, and RIE treatments on properties of double-sided high voltage solar cells with vertically aligned p-n junctions, Int. J. Photoenergy **2016**, 1815205 (2016). https://doi.org/10.1155/2016/1815205.

24. A. A. Evtukh, V. G. Litovchenko, N. I. Klyui, M. O. Semenenko, E. B. Kaganovich, E. G. Manoilov, Electron field emission from porous silicon prepared at low anodisation currents, Int. J. Nanotechnol. **3**, 89 (2006). https://doi.org/10.1504/IJNT.2006.008723.

25. M. Semenenko, S. Antonin, R. Redko, Yu. Romanuyk, A. V. Hladkovska, V. Solntsev, A. Evtukh, Resonant tunneling field emission of Si sponge-like structures, J. Appl. Phys. **128**, 114302 (2020). https://doi.org/10.1063/5.0020527.

26. N. A. Hill, K. Birgitta Whaley, A theoretical study of light emission from nanoscale silicon, J. Electron. Mater. **25**, 1132 (1996). https://doi.org/10.1007/BF02659915.

27. H. Yorikawa, S. Muramatsu, Logarithmic normal distribution of particle size from a luminescence line-lineshape, Appl. Phys. Lett. **71**, 644 (1997). https://doi.org/10.1063/1.119816.